# Deep Coherence Learning: An Unsupervised Deep Beamformer for High Quality Single Plane Wave Imaging in Medical Ultrasound

Hyunwoo Cho[1], Seongjun Park[1], Jinbum Kang[2,*] and Yangmo Yoo[1,3*]

[1]Department of Electronic Engineering, Sogang University, Seoul 04107, Korea

[2]Department of Bioengineering, University of Washington, Seattle, WA 98105, United States

[3]Department of Biomedical Engineering, Sogang University, Seoul 04107, Korea

*Corresponding Authors: jkang7@uw.edu; ymyoo@sogang.ac.kr

**Abstract**

Plane wave imaging (PWI) in medical ultrasound is becoming an important reconstruction method with high frame rates and new clinical applications. Recently, single PWI based on deep learning (DL) has been studied to overcome lowered frame rates of traditional PWI with multiple PW transmissions. However, due to the lack of appropriate ground truth images, DL-based PWI still remains challenging for performance improvements. To address this issue, in this paper, we propose a new unsupervised learning approach, i.e., deep coherence learning (DCL)-based DL beamformer (DL-DCL), for high-quality single PWI. In DL-DCL, the DL network is trained to predict highly correlated signals with a unique loss function from a set of PW data, and the trained DL model encourages high-quality PWI from low-quality single PW data. In addition, the DL-DCL framework based on complex baseband signals enables a universal beamformer. To assess the performance of DL-DCL, simulation, phantom and *in vivo* studies were conducted with public datasets, and it was compared with traditional beamformers (i.e., DAS with 75-PWs and DMAS with 1-PW) and other DL-based methods (i.e., supervised learning approach with 1-PW and generative adversarial network (GAN) with 1-PW). From the experiments, the proposed DL-DCL showed comparable results with DMAS with 1-PW and DAS with 75-PWs in spatial resolution, and it outperformed all comparison methods in contrast resolution. These results demonstrated that the proposed unsupervised learning approach can address the inherent limitations of traditional PWIs based on DL, and it also showed great potential in clinical settings with minimal artifacts.

# 1. Introduction

Medical ultrasound is a useful clinical imaging modality with the advantages of real-time imaging, no ionizing radiation and relatively low cost. Plane wave imaging (PWI) has become an important reconstruction method with high frame rates, and it facilitates a variety of new clinical applications, such as shear wave elastography and microvascular imaging (Bercoff et al., 2011; Taljanovic et al., 2017; Tanter and Fink, 2014). For PWI, a single PW transmission and reconstruction (i.e., single PWI) would be the most favorable approach (i.e., one firing for a final single frame), but the image quality is unfortunately very poor (Hendriks et al., 2022; Montaldo et al., 2009). To increase image quality, such as spatial resolution and contrast, in single PWI, a coherent PW compounding technique (CPWC), which transmits a set of multiple PWs with a range of steering angles, was introduced and has become an essential technique for high-quality PWI (Hendriks et al., 2022; Montaldo et al., 2009). However, several PW transmissions decrease the frame rate by a factor of the number of PWs, and it eventually allows a limited number of PW transmissions to maintain high temporal resolution. Therefore, the improvement of image quality with limited transmissions is still challenging in PWI.

To enhance the image quality without sacrificing the frame rate, various adaptive beamforming methods have been introduced for PWI (Austeng et al., 2011; Nguyen and Prager, 2018; Wang et al., 2020; Yang et al., 2020). One of the representative approaches is to utilize weighting factors such as the coherence factor (CF) (Hollman et al., 1999; Li and Li, 2003). For example, the calculated weighting factor to suppress sidelobes in a beam pattern is applied between multiple PW frames in a set of CPWCs, and it improves contrast while effectively reducing the reflected noise and artifacts (Wang et al., 2019). However, the methods based on such weighting factors generate different types of artifacts near hyperechoic signals, i.e., dark region artifacts, and they may have the potential to distort clinical information. Another approach for PWI is to apply the minimum-variance (MV) beamformer, which employs the statistical properties of radiofrequency (RF) signals to calculate weight vectors, but the computational cost is extremely high while obtaining higher resolution and contrast (Austeng et al., 2011; Synnevag et al., 2007). Recently, PWI with the delay-multiply-and-sum (DMAS) beamforming technique, which enhances signal coherence using adjacent multiple channel signals, has been introduced along with various modifications, and it has shown high potential in clinical settings with diverse applications (Kang et al., 2020; Matrone et al., 2016; Shen and Chu, 2021; Shen and Hsieh, 2019; Shen and Tu, 2020). However, the nonlinearity of multiplication operations can be significant in signal distortion, and the computational complexity also remains challenging.

Recently, artificial intelligence (AI)-driven ultrasound imaging techniques have been widely proposed to solve tasks such as image classification, segmentation, and reconstruction (Akkus et al., 2019; Micucci and Iula, 2022; Wang et al., 2021). Specifically, deep learning (DL)-based adaptive beamforming and reconstruction methods have been studied for PWI and its applications (Bell et al., 2020; Liebgott et al., 2016; Wang et al., 2021). For instance, a generative adversarial network (GAN) was adopted and trained as an alternative to traditional delay-and-sum (DAS) beamforming in ultrasound imaging (Nair et al., 2019). In addition, the GAN architecture was utilized to transform low image quality of PWI into higher image quality, such as conventional focused beam imaging (Zhou et al., 2019). For this, a unique data acquisition scheme was designed to simultaneously obtain PW images and focused beam images for input and ground-truth data, respectively. Another suggested approach was to exploit deep neural networks to calculate an optimal set of apodization weights for adaptive beamforming, and it also showed strong benefits in terms of data efficiency with enhanced image quality (Luijten et al., 2020). On the other hand, several convolutional neural network (CNN)-based beamforming techniques (e.g., modified U-Net (Nguon et al., 2022) and cascaded CNN architecture (Wasih et al., 2023)) have been developed for PWI, and high-quality ground truth data were generated by CPWC with many PW frames (Nguon et al., 2022;

Tang et al., 2021; Wasih et al., 2023).

Studies of DL-based beamforming networks based on supervised learning demonstrated that they can significantly improve the image quality of PWI along with several benefits. However, they still have some limitations deriving from the nature of supervised learning that relies on ground-truth data. First, the generation of ground-truth data for training may be problematic due to the lack of reference images in medical ultrasound (Zhang et al., 2021), and the inability to generalize to a variety of applications. For example, a simulated dataset for ground truth is very likely to train the networks to be valid only for simulation images (Goudarzi and Rivaz, 2022; Nair et al., 2019), and it may cause additional stages to validate its performance on real *in vivo* images. In addition, specially designed data acquisition methods such as interleaving (Zhou et al., 2019) may be inappropriate for cost-effective ultrasound systems. Second, the performance of supervised learning approaches relies on the image quality of ground-truth data since the networks are trained to mimic the ground truth using certain loss functions (Tang et al., 2021; Wasih et al., 2023). To address these limitations, a self-supervised learning framework was recently adopted for PWI where RF data are employed for both inputs and labels during DL training (Zhang et al., 2021). However, establishing a particular measurement matrix for regularization is still a challenge; the method may be only effective for certain situations, e.g., certain imaging probes, based on the forward model.

Unsupervised learning can potentially address the constraints of supervised learning approaches for PWI tasks. Since the unsupervised learning approach is not allowable for ground-truth data, it requires a unique training strategy to optimize the DL networks. In this paper, we propose a new unsupervised learning approach, i.e., deep coherent learning (DCL), that leverages the signal characteristics of PWI from a set of CPWC data in the training scheme of the DL network. By using the proposed method, the DL network is not limited to the quality of ground-truth data and can be more reliable in various circumstances without reference images. The main contributions of this paper can be summarized as follows: (1) We propose a novel training strategy and a loss function that exploits the signal coherence between different PWs in a set of CPWC to facilitate an unsupervised learning approach for DL-based PWI. The training framework enables the DL network to learn the signal coherence from low-quality data pairs; it can address concerns about overfitting issues or performance constraints. Therefore, the DL model can reconstruct high-quality PWI images from single plane wave data. (2) The proposed training strategy trains DL networks as a universal beamformer for PWI since the DL network employs complex signals (i.e., in-phase and quadrature (I/Q) signals) as its inputs and outputs before postprocessing for final image reconstruction. Thus, it allows us to transform the proposed DL framework into a universal beamformer that can be implemented in various circumstances (e.g., different types of transducers or systems). (3) To evaluate the efficacy of the proposed method in a more reasonable manner, publicly available *in vivo* and phantom datasets were compared and analyzed.

## 2. Related Works

Single PWI can produce an image frame with an exceptionally high frame rate compared to conventional focused beam imaging or CPWC imaging. However, the image quality of a single PWI is extremely limited due to only one-way dynamic beam focusing in reception (Montaldo et al., 2009). Thus, it suffers from several artifacts caused by relatively higher axial, side and grating lobes in the beam pattern; the contrast resolution is severely degraded. Recent studies have shown the great potential of DL networks to improve image quality by suppressing artifacts in a single PWI (Li et al., 2020; Lu et al., 2022; Nguon et al., 2022; Wasih et al., 2023). Li et al. proposed a U-Net-based encoder-decoder architecture, and CPWC (i.e., 75-plane waves) was employed to generate ground-truth images for a supervised learning framework (Li et al., 2020). However, the feasibility study showed limitations in terms of spatial resolution and blurred features while effectively suppressing speckle noise and artifacts. Lu et al. also adopted the U-Net architecture, which consists of a two-stage training strategy (Lu et al., 2022). In the first pretraining, simulation and phantom datasets were acquired to train the CNN model, and additional phantom and *in vivo* experimental datasets were utilized for the second transfer learning. The U-Net-based CNN architecture showed that it can enhance the image quality of a single PWI in terms of spatial resolution and contrast-to-noise ratio (CNR). However, it revealed a typically limited aspect of supervised learning frameworks since the performance was constrained by the ground-truth data using 3- or 5- PWs of CPWC. In addition, another U-Net-based DL network architecture was introduced for a single PWI (Nguon et al., 2022), and different networks (i.e., EfficientNet (Tan and Le, 2019) and U-Net (Ronneberger et al., 2015)) were utilized for each encoder and decoder stage for DL-based beamforming. The results showed that the proposed architecture can outperform the original U-Net and EfficientNet models. More recently, Wasih et al. proposed a cascaded DL network combined with a fully CNN and a conditional GAN to improve the performance of a single PWI (Wasih et al., 2023). Although these studies demonstrated that diverse structures of DL networks can help to improve the image quality of a single PWI by outperforming typical neural networks, the imaging performance is still limited by the ground-truth data that are generated by a limited number of PWs for CPWCs.

Supervised learning-based DL network approaches have demonstrated improvements in image quality, such as CNR. However, the lack of an available ideal ground-truth dataset can still limit the best performance in supervised learning frameworks, and it is also challenging to find a reasonable acquisition method for ideal datasets. To address these constraints, a self-supervised learning approach was proposed for single PWI and CPWC (Zhang et al., 2021); a DL network trained with a loss function including sparse regularization parameters was presented, and RF data were employed for both the inputs and labels during training in a self-supervised manner. However, the precomputed measurement matrix for regularization is probe-dependent due to the forward model, and distorted images by severe dark region artifacts may not be useful in clinical settings.

## 3. Material and methods

*3.1. Deep Coherence Learning: Unsupervised Deep Beamformer for a Single PWI*

*3.1.1. Principle and Framework*

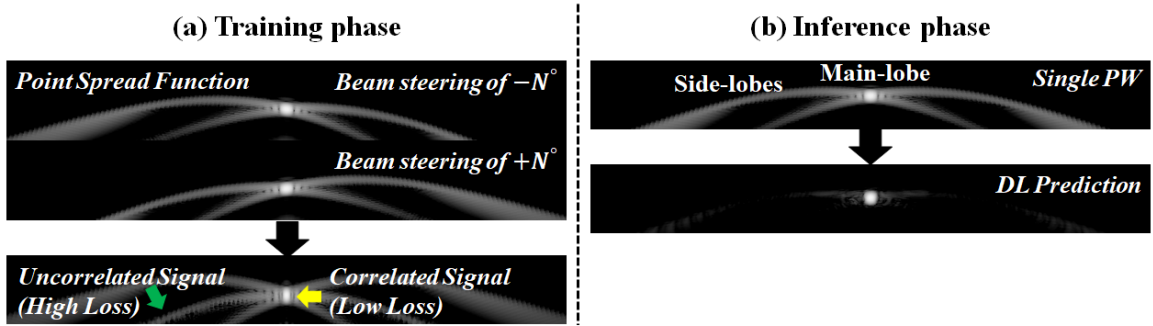

**Fig. 1.** Illustrations of the fundamental principle of deep coherence learning (DCL) from two different PWs (i.e., steering angles with $-N°$ and $+N°$) as an example. In the training phase, as shown in (a), different losses were calculated by using signal coherence between the two point spread functions (PSFs). Then, a DL model is trained to suppress uncorrelated signals (e.g., side lobes) by a loss function, and the trained DL model predicts only highly correlated signals from input data (i.e., a single plane wave), as shown in (b).

To address the limitations of supervised or self-supervised learning approaches, a new unsupervised learning approach, called deep coherence learning (DCL), is proposed for a single PWI. Instead of training DL networks with the referenced data (i.e., ground truth), the signal coherence between multiple PWs in the set of CPWC is identified and enforced during training of DL networks in an unsupervised manner. Fig. 1 describes the fundamental principle of the DCL from the only two different PWs with a range of $-N°$ to $+N°$ as an example. As illustrated in Fig. 1(a), highly correlated signals between the two PWs, e.g., main lobes in point spread functions (PSFs), are enhanced with relatively low loss by a unique loss function, while uncorrelated signals, e.g., side lobes or grating lobes in PSFs, are suppressed with relatively higher loss during the training phase. Thus, DCL leads the DL model to predict highly coherent signals from several single PWs in the CPWC dataset, and the trained DL model encourages a high-quality PWI from low-quality single PW data, as shown in Fig. 1(b).

Fig. 2 illustrates the unsupervised learning framework based on DCL with a dataset of CPWC and the single PWI using the trained DL network. In the training phase, as illustrated in Fig. 2(a), complex signals (i.e., in-phase and quadrature (I/Q) signals) with a set of multiple PWs (i.e., P = $\{P_1, P_2, ..., P_k\}$, $k$ = the number of PW transmissions with a range of $-N°$ to $+N°$) are initially produced after receive beamforming (e.g., delay-and-sum, DAS (Steinberg, 1992)). Then, each I/Q data frame for a set of PWs is used differently for training the DL network model in accordance with an unsupervised training strategy. For this, one of the PW I/Q datasets (P) is randomly sampled from the PW I/Q dataset (P) as the input data ($P_i$), and it is immediately fed into the DL network. The DL network is then trained by computing a loss between the prediction and the remaining frames (i.e., $\{P_1, P_2, ..., P_{k \neq i,v}\}$) except for the validation data ($P_v$), as illustrated in Fig. 2(a). After the training phase, the trained DL network using DCL is utilized as a beamforming module (i.e., DCL beamformer), where the input and output are defined as a single I/Q frame, as shown in Fig. 2(b), and it can be expressed by:

$$y_{DAS}(t) = \frac{1}{N}\sum_{i=1}^{N} w_r(i)x_i(t - \tau_i) \tag{1}$$

$$y_{DL}(t) = f(y_{DAS}(t)) \tag{2}$$

where $y_{DAS}$ and $y_{DL}$ are the output of DAS beamforming for time delay compensation and the output of the DCL beamformer for adaptive beamforming, respectively. $f$ represents the DCL beamformer including I/Q signal conversion (e.g., Hilbert transform) for RF signals, $w_r$ indicates the apodization weights of the $i_{th}$ element, $x_i$ is the received RF data, and $\tau_i$ denotes the time of flight (TOF). After DCL beamforming, postprocessing, such as envelope detection and log compression, for B-mode images is performed using DL prediction ($y_{DL}$), and a high-quality single PW image is finally obtained, as described in Fig. 2(b).

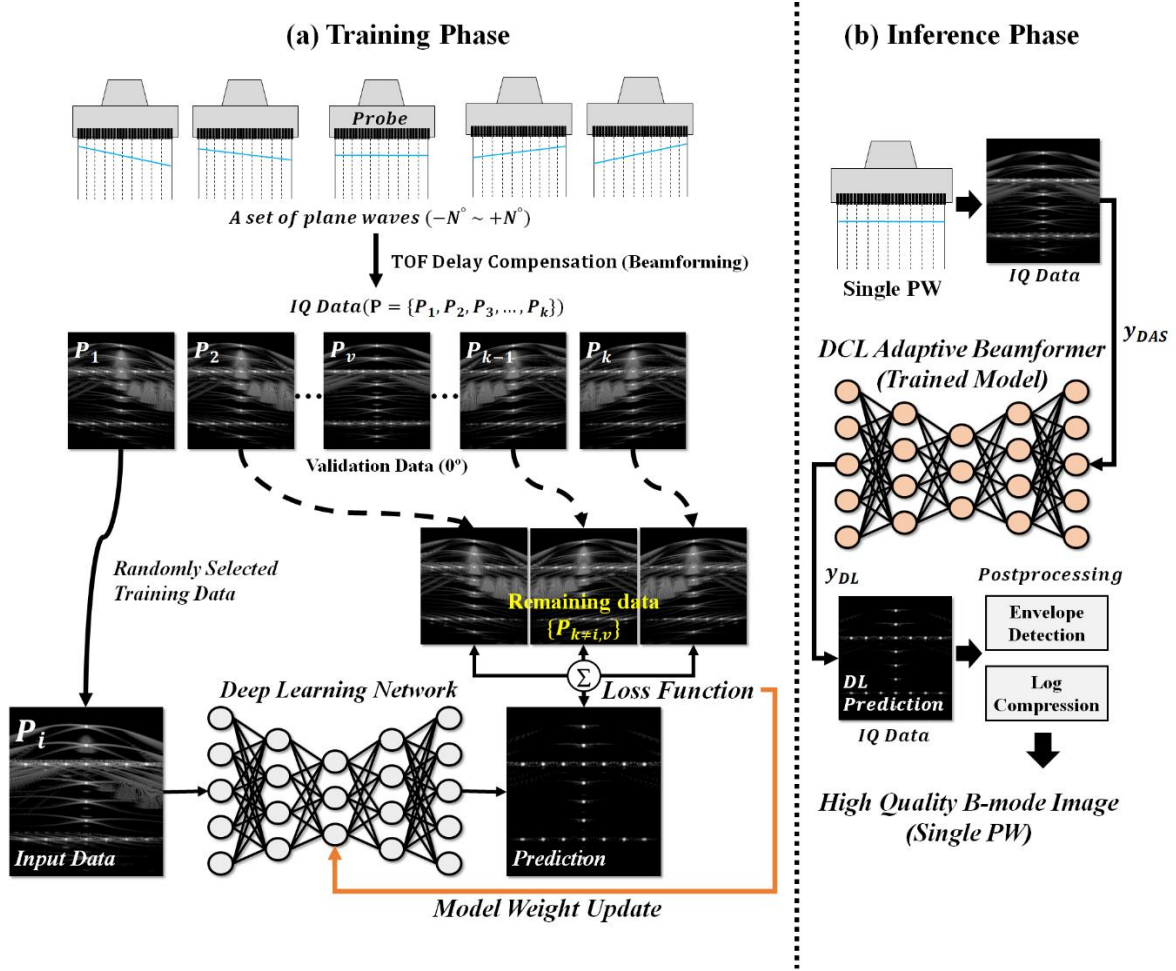

**Fig. 2.** Schematic representation of the proposed DCL framework. (a) In the training phase, an RF dataset of CPWC (i.e., a set of multiple PWs with different steering angles in a range of $-N°$ to $+N°$) is beamformed by the typical delay-and-sum (DAS) approach, and complex signals (i.e., I/Q data) are produced (i.e., $\mathbf{P} = \{P_1, P_2, ..., P_k\}$). Then, the input data ($P_i$) are randomly selected among the I/Q dataset (P) and fed into the DL network. The network is trained by calculating a loss between the prediction and the remaining frames (i.e., $\{P_1, P_2, ..., P_{k \neq i,v}\}$), which excludes the validation data ($P_v$). (b) In the inference phase, single PW data (i.e., I/Q signals) are fed into the trained network (i.e., DCL beamformer), and the predicted output data (i.e., I/Q signals) are passed through postprocessing (e.g., envelope detection and log compression) to reconstruct a final B-mode image.

*3.1.2. Coherence Loss*

To quantify the difference between two statistically independent signals (e.g., additive white Gaussian noise), loss functions such as the mean squared error (MSE) (Lehtinen et al., 2018) or extended Stein's unbiased risk estimate (eSURE) (Zhussip et al., 2019) have typically been employed for unsupervised learning tasks. However, their use for identifying signal coherence from severe artifacts of PW images might be limited. For this reason, a coherence loss function was developed to quantify the correlation between predicted and target signals for DCL. The coherence loss ($L_{Coherence}$) measures the correlation between the prediction (i.e., $f(P_i)$) and target data ($P_t$) for every pair of the input PW dataset:

$$L_{Coherence} = \sum_{t=1}^{k} \left(-\frac{f(P_i) \times P_t^*}{\sqrt{f(P_i) \times f(P_i)^*} \times \sqrt{P_t \times P_t^*}}\right) (t \neq i) \tag{3}$$

where $P_i$ and $P_t$ denote the input and target I/Q data, respectively, and $f$ represents the deep learning prediction. In addition, $*$ is the complex conjugate of I/Q signals, and $k$ is the total number of PW images of CPWC. A negative sign of $L_{Coherence}$ contributes to maximizing highly correlated signals as well as minimizing the overall loss.

*3.2. Network Architecture and Implementation Details*

To validate the proposed training strategy with the coherence loss function, a convolutional neural network (CNN) based on the U-Net architecture (Ronneberger et al., 2015) was employed. As depicted in Fig. 3, the dimension for the input and output data of the network was 2×256×256, which is determined as the two randomly cropped images (i.e., I and Q data) with the size of 256×256. Based on the encoding and decoding structures, multiple convolution blocks, which incorporate two $3 \times 3$ convolution layers and the LeakyReLU (Maas et al., 2013) activation function, were constructed. To preserve the negative components of signals, LeakyReLU was chosen as the activation function for each convolution block. In the encoding stage, downsampling is executed as the number of filters gradually increases. Conversely, in the decoding stage, the extracted features are gradually upsampled to their original dimensions, incorporating skip connections between the encoder and decoder layers. The network output (i.e., prediction) is constrained within the range of (−1, 1) using the tanh activation function. Training of the network is performed using the AdamW (Loshchilov and Hutter, 2017) optimizer with a cosine annealing learning rate scheduling (Loshchilov and Hutter, 2016) that employs a continuously changing learning rate. The initial learning rate was $1e^{-4}$ and scheduled in the range of minimum $1e^{-7}$ with a period of 20,000 steps.

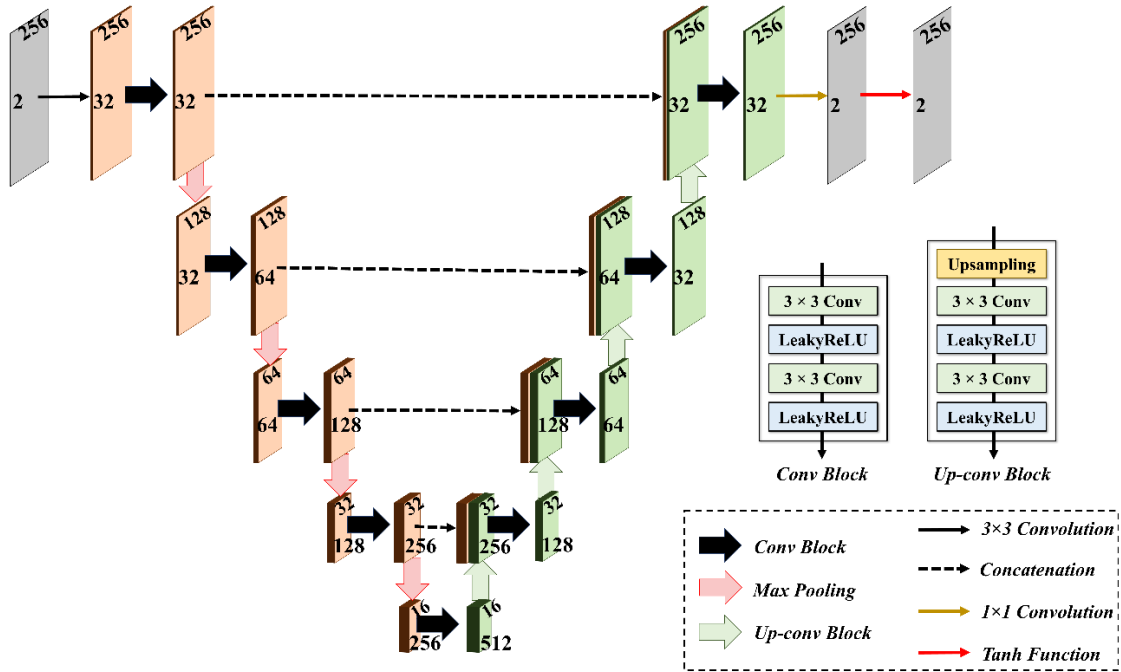

**Fig. 3.** The proposed CNN architecture for DCL training. The network comprises multiple convolution blocks and the tanh activation function in the encoding and decoding stages. Each convolution block consists of two 3×3 convolutions and LeakyReLU activation functions. The number of filters in the convolution blocks gradually increases (i.e., 32, 64, 128, 256, and 512) with the depth of the network. In addition, skip connections are conducted between the encoding and decoding layers to retain essential features.

*3.3. Experimental Setup and Evaluation Metrics*

To train and evaluate the proposed DCL beamformer, two public datasets, i.e., CUBDL (Hyun et al., 2021) and PICMUS (Liebgott et al., 2016), were utilized. The CUBDL (Hyun et al., 2021) dataset is a large-scale public database that includes various subdatasets acquired by different research groups under various circumstances (e.g., ultrasound machines, transducers, targets). The entire database comprises 576 image acquisition sequences. The PICMUS (Liebgott et al., 2016) dataset is another public dataset that includes simulation and experimental phantom data, as well as two *in vivo* datasets. Among varied image acquisition sequences from the two datasets, 5 of MYO, 5 of INS, 2 of UFL, 11 of JHU (Li et al., 2020) and 50 of TSH (Zhang et al., 2018) were employed from the CUBDL, and 6 of PICMUS were also utilized. The total number of frames was 3715 that is composed of 79 validation frames and other training frames. Therefore, this study included 3 different types of ultrasound machines and 7 different types of transducers for generalization. In addition, simulation, phantom, and *in vivo* data were jointly employed for the trained model. Training of the network was implemented using RTX3090 (NVIDIA, Santa Clara, CA, USA) with the PyTorch framework. The total training time was approximately 3 GPU days.

For comparison with several beamforming methods, conventional delay-and-sum (DAS) (Montaldo et al., 2009) and delay-multiply-and-sum (DMAS) (Matrone et al., 2016) were implemented as traditional approaches. For DAS beamforming, a single PW and 75-PWs were exploited to compare the low and high image quality of PWI at the same time. DMAS adaptive beamforming was applied to single PW data, and out-of-band signals other than doubled frequency components were properly cut off (i.e., f-DMAS (Matrone et al., 2014))). In addition to the rule-based traditional beamforming algorithms, other DL-based beamforming models, i.e., DL beamformer with supervised learning (DL-SP) and DL beamformer with GAN (DL-GAN), were also compared. To implement the DL-SP, the proposed CNN network (Fig. 3) was utilized in the same manner, and it was trained by the MSE loss function with the best quality ground-truth data (i.e., 75-PWs). For DL-GAN, the generator was implemented using the same CNN network (Fig. 3), and only the encoder part of the proposed network was employed for the discriminator. All DL models were trained with the same number of steps for a fair comparison, and the best results were selected for validation.

To evaluate the imaging performance, spatial resolutions (i.e., full width at half maximum (FWHM)) were axially and laterally measured, and axial artifacts caused by axial lobes of PSFs in PWI were also measured by the peak range axial-lobe level (PRAL). In addition, both the contrast-to-noise ratio (CNR) (Chow and Paramesran, 2016) and the generalized contrast-to-noise ratio (gCNR) (Khan et al., 2020; Rodriguez-Molares et al., 2019; Wiacek et al., 2020) were measured to assess image contrast. The CNR can be calculated as:

$$\text{CNR (dB)} = 20 \log_{10} \frac{|\mu_i - \mu_o|}{\sqrt{\sigma_i^2 + \sigma_o^2}} \quad (4)$$

where $\mu$ and $\sigma$ represent the mean and standard deviation of isoechoic (i.e., $\mu_i$ and $\sigma_i$) and anechoic (i.e., $\mu_o$ and $\sigma_o$) regions, respectively. The gCNR, which is a relatively new image quality metric (Rodriguez-Molares et al., 2019), was measured by:

$$\text{gCNR} = 1 - \int min\{d_i(x), d_o(x)\}dx \quad (5)$$

where $d_i$ and $d_o$ represent the probability density functions of the isoechoic and anechoic regions, respectively.

## 4. Results

*4.1. Simulation Study*

*4.1.1. Point-target phantom*

To evaluate the spatial resolution and the levels of secondary lobes of a single PWI, a simulation study using point targets was conducted. Fig. 4 presents the reconstructed B-mode results (a dynamic range of 60 dB) of a single PWI using the two traditional beamformers (i.e., DAS and DMAS) and the three DL-based beamformers (i.e., DL-SP, DL-GAN and the proposed DL-DCL) from the simulated point-target phantom. The DAS with 75-PWs (i.e., CPWC) was also used as a reference (ground truth), and it showed the best image quality while significantly suppressing all PW artifacts (i.e., grating lobes, side lobes and axial lobes), as illustrated in Fig. 4(b). Fig. 4(c) presents single PWI with DMAS adaptive beamforming, and it effectively suppressed grating lobe and side lobe artifacts, while relatively high axial lobe artifacts remained. The two representative DL models based on supervised learning and GAN architecture (i.e., DL-SP and DL-GAN) produced better image quality than conventional single PWI (i.e., DAS (1-PW)) by suppressing overall artifacts; however, the level of suppression in artifacts was worse than DMAS (1-PW), as illustrated in Figs. 4(d) and 4(e). Fig. 4(f) illustrates the proposed DL method based on the unsupervised learning framework (i.e., DCL), and it outperformed the other two DL-based methods (i.e., DL-SP and DL-GAN) and DMAS in terms of artifact suppression. However, the low level of artifacts remained in the near depth (i.e., 5 to 10 mm) due to the limited field-of-view of PWI (i.e., the signals in the near depth are less correlated due to the limited transmit synthetic focusing for steered PW beams (Kang et al., 2019)).

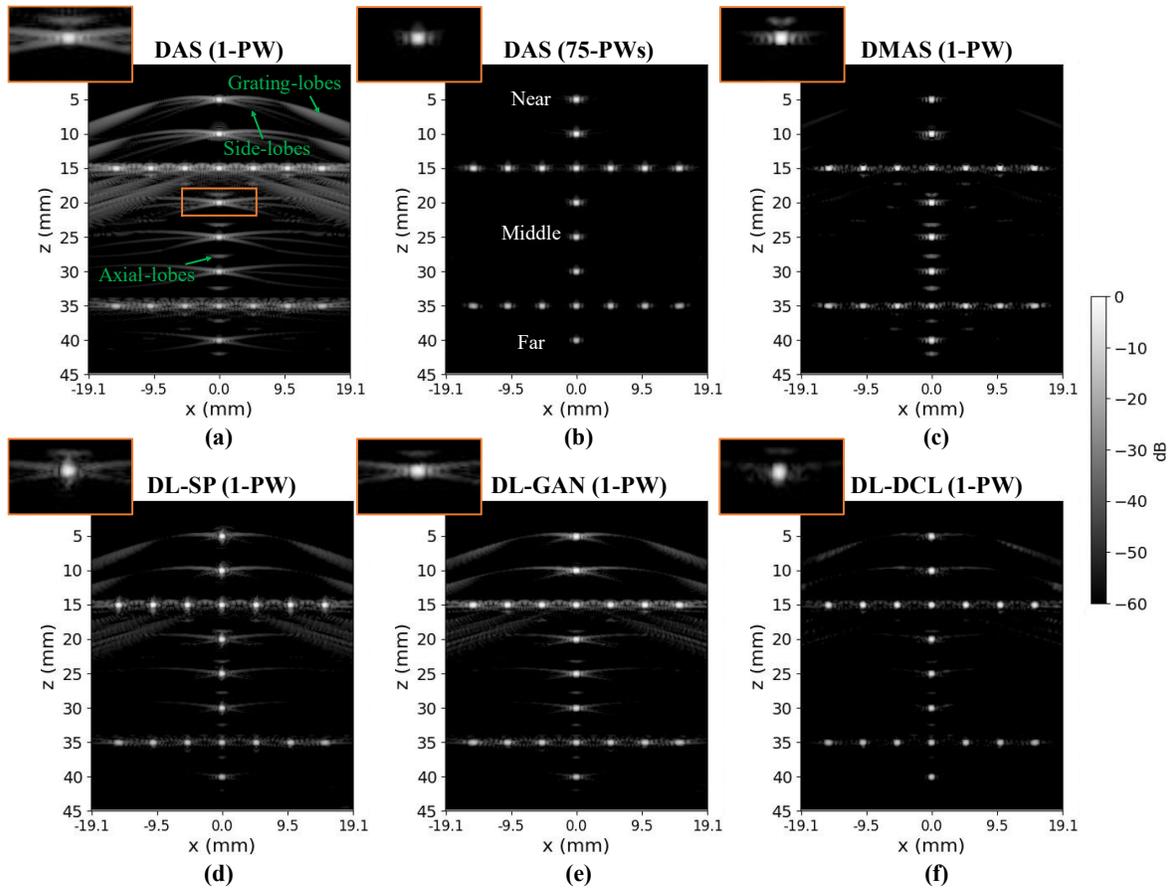

**Fig. 4.** Simulated B-mode images for the point-target phantom reconstructed by (a) DAS with 1-PW, (b) DAS with 75-PWs, (c) DMAS with 1-PW, (d) supervised learning-based DL method with 1-PW, (e) GAN-based DL method with 1-PW and (f) the proposed DCL-based DL method with 1-PW. The green arrows in (a) indicate the three different types of artifacts caused by secondary lobes (i.e., side lobes, grating lobes and axial lobes) from conventional DAS beamforming in single PWI.

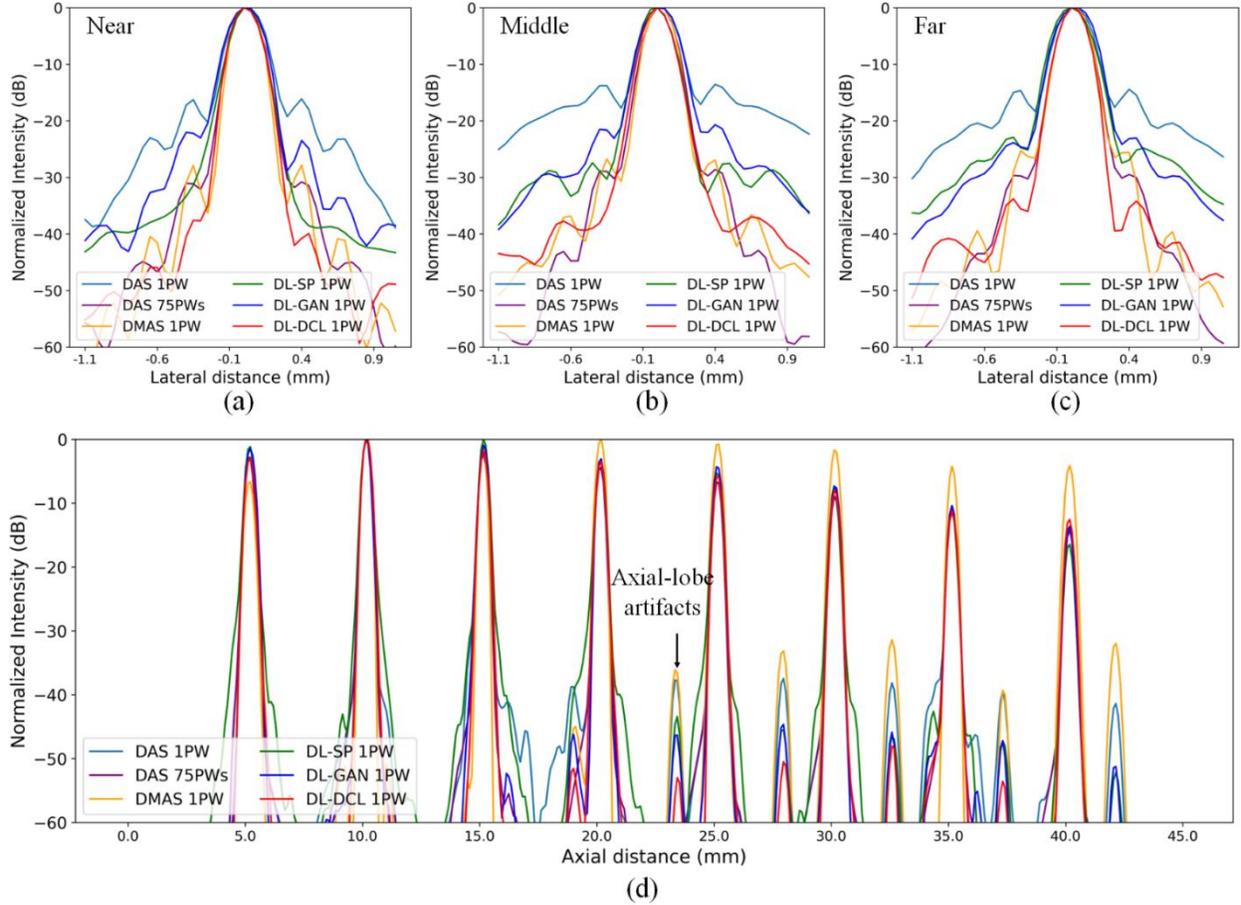

**Fig. 5.** (a)-(c) Normalized lateral profiles of the six beamforming methods (Fig. 4) in the point targets corresponding to the near, middle and far depths (i.e., 5, 25, and 40 mm) indicated by Fig. 4(b). (d) Normalized axial profiles of the six beamforming methods (Fig. 4) in the center scanline region ($x = 0.0$ mm).

To quantitatively assess the performance of the six beamforming methods, lateral and axial FWHMs and PRAL were measured based on each lateral and axial profile. Fig. 5 illustrates the normalized lateral and axial profiles for the point targets corresponding to three different depths (i.e., near, middle and far depths indicated in Fig. 4(b)) and the center scanline ($x = 0$ mm), and their measurements are listed in Tables 1 to 3. As shown in Figs. 5(a)-5(c) and Table 1, conventional DAS with 1-PW produced the highest side-lobe level and the widest main-lobe width at all depths, whereas DAS with 75-PWs and DMAS with 1-PW yielded relatively low side-lobe levels and narrow main-lobe widths. The two DL-based methods (i.e., DL-SP and DL-GAN) mostly showed higher side-lobe levels and wider main-lobe widths than DMAS, as illustrated in Figs. 5(a)-5(c) and as listed in Table 1. Compared to the two DL-based methods, the proposed DL-DCL method with 1-PW produced the lowest levels of side lobes and the narrowest main-lobe widths among the six images. For the axial profiles, the six beamforming methods showed similar main-lobe widths for most targets, as illustrated in Fig. 5(d). However, the four-comparison methods excluding DAS with 75-PWs (i.e., DAS with 1-PW, DMAS with 1-PW, DL-SP with 1-PW, DL-GAN with 1-PW) suffered from high levels of axial lobe artifacts, while DL-DCL with 1-PW produced the lowest level among all methods excluding DAS with 75-PWs, as shown in Table 3.

**Table 1.** The measured lateral FWHMs (mm) for the three-point targets corresponding to the near, middle and far depths (5, 25 and 40 mm) in the six beamforming methods.

|       | DAS (1-PW) | DAS (75-PWs) | DMAS (1-PW) | DL-SP (1-PW) | DL-GAN (1-PW) | DL-DCL (1-PW) |
|-------|------------|--------------|-------------|--------------|---------------|---------------|
| 5 mm  | 0.49       | 0.39         | 0.30        | 0.39         | 0.49          | 0.39          |
| 25 mm | 0.49       | 0.39         | 0.30        | 0.39         | 0.49          | 0.39          |
| 40 mm | 0.49       | 0.39         | 0.39        | 0.49         | 0.49          | 0.39          |

**Table 2.** The measured axial FWHMs (mm) for the three-point targets corresponding to the near, middle and far depths (5, 25 and 40 mm) in the six beamforming methods.

|       | DAS (1-PW) | DAS (75-PWs) | DMAS (1-PW) | DL-SP (1-PW) | DL-GAN (1-PW) | DL-DCL (1-PW) |
|-------|------------|--------------|-------------|--------------|---------------|---------------|
| 5 mm  | 0.30       | 0.30         | 0.30        | 0.30         | 0.39          | 0.39          |
| 25 mm | 0.30       | 0.30         | 0.30        | 0.30         | 0.30          | 0.30          |
| 40 mm | 0.39       | 0.39         | 0.39        | 0.30         | 0.30          | 0.39          |

**Table 3.** The measured PRALs (dB) for the five-point targets (20, 25, 30, 35 and 40 mm) at the center scanline in the six beamforming methods.

|       | DAS (1-PW) | DAS (75-PWs) | DMAS (1-PW) | DL-SP (1-PW) | DL-GAN (1-PW) | DL-DCL (1-PW) |
|-------|------------|--------------|-------------|--------------|---------------|---------------|
| 20 mm | 33.37      | 66.70        | 36.10       | 41.15        | 43.35         | 52.50         |
| 25 mm | 30.81 | 65.17 | 32.37 | 40.18 | 40.35 | 44.60 |
| 30 mm | 29.28 | 63.15 | 29.71 | 39.18 | 38.51 | 40.04 |
| 35 mm | 28.28 | 56.07 | 35.07 | 36.04 | 36.81 | 42.34 |
| 40 mm | 27.26 | 55.24 | 27.86 | 35.72 | 37.61 | 47.66 |

*4.1.2. Anechoic cyst phantom*

Fig. 6 illustrates the reconstructed B-mode images (a dynamic range of 60 dB) using the six beamforming methods (i.e., DAS with 1-PW, DAS with 75-PWs, DMAS, DL-SP with 1-PW, DL-GAN with 1-PW and the proposed DL-DCL with 1-PW) for the anechoic cyst phantom. As shown in Figs. 6(a) and 6(b), the traditional DAS beamforming method with 1-PW suffers from low-contrast resolution caused by high side-lobe artifacts, whereas DAS with 75-PWs yielded much higher contrast resolution with clear cystic targets. DMAS adaptive beamforming with 1-PW showed low image contrast despite lower side-lobe levels since the nonlinear property gave rise to signal distortion (i.e., dark region artifacts (Rindal et al., 2017)) and resulted in reduced dynamic ranges (i.e., black and white images), as depicted in Fig. 6(c). In addition, as illustrated in Figs. 6(d) and 6(e), the two DL model-based approaches (i.e., DL-SP and DL-GAN) with 1-PW produced better image quality (i.e., higher contrast) than DAS with 1-PW, but they still suffer from noisy cystic targets and unclear boundary representations due to high PW artifacts. However, the proposed unsupervised learning-based DL method (i.e., DL-DCL) yielded higher image contrast for all cystic targets as well as clear boundaries without dark region artifacts.

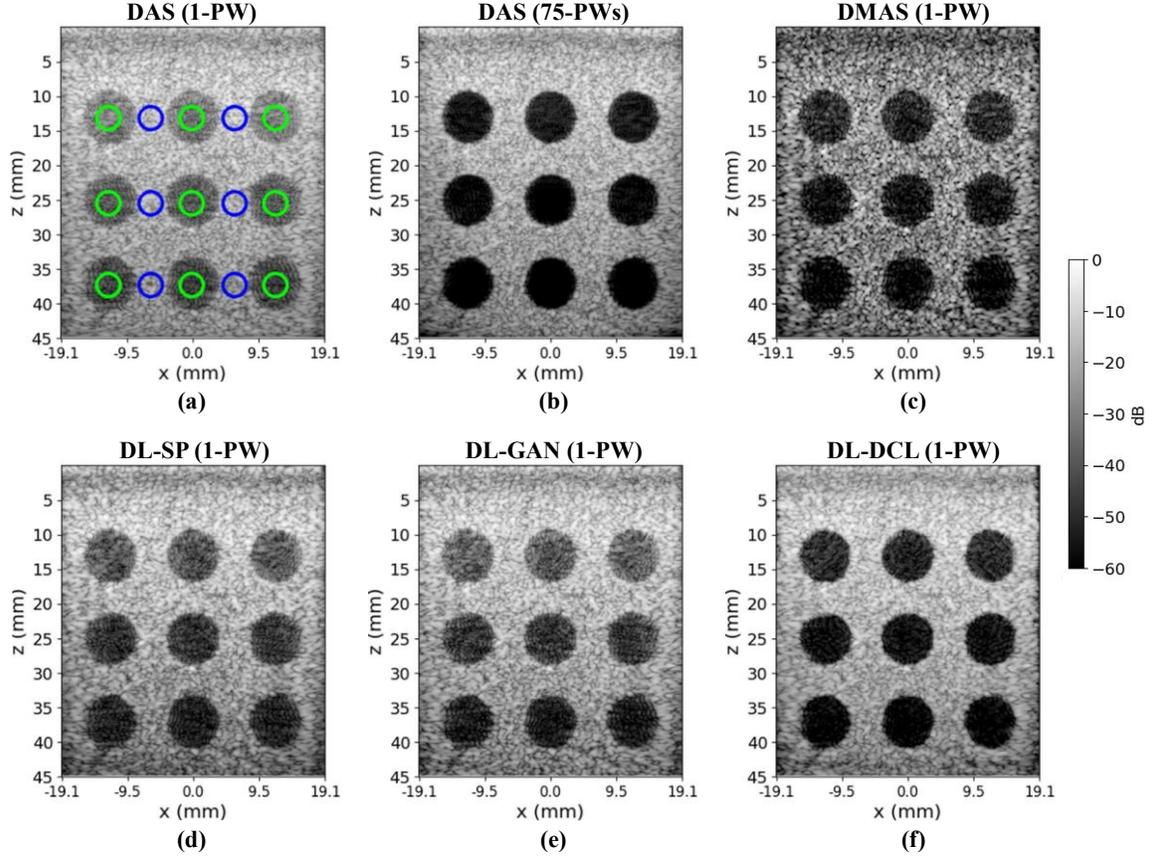

**Fig. 6.** Simulated B-mode images for the cystic phantom (9-anechoic targets) reconstructed by (a) DAS with 1-PW, (b) DAS with 75-PWs, (c) DMAS with 1-PW, (d) DL-SP with 1-PW, (e) DL-GAN with 1-PW and (f) the proposed DL-DCL with 1-PW. The green and blue circles in (a) indicate the anechoic and isoechoic regions, respectively, for contrast measurements (i.e., CNR and gCNR).

Fig. 7 depicts the box plots of each CNR and gCNR measurement (i.e., Eqs. (4) and (5)) for 9-anechoic targets, as indicated by Fig. 6(a). As illustrated in Fig. 7, DAS with 1 PW showed the lowest CNR and gCNR values, while DAS with 75-PWs produced the highest CNR and gCNR values, similar to the visual assessment (i.e., Fig. 6). Unlike the point-target study (Fig. 4), the two DL-based methods (i.e., DL-SP and DL-GAN) outperformed DMAS for both CNR and gCNR measurements since severe dark regions of DMAS increased speckle variance on isoechoic regions (i.e., background). On the other hand, the proposed DL-DCL method showed comparable results with DAS with 75 PWs in terms of both CNR and gCNR, i.e., 17.6 vs. 18.3 and 0.99 vs. 0.99, respectively, as depicted in Figs 7(a) and 7(b).

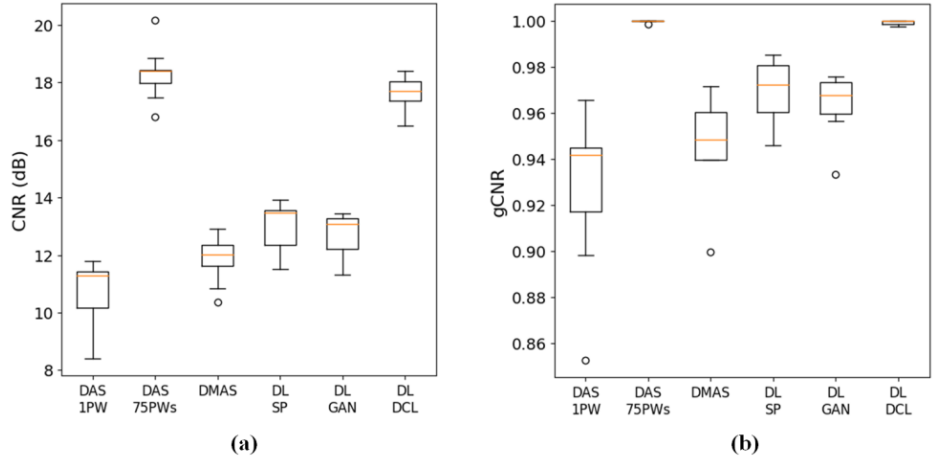

**Fig. 7.** Box plots for (a) CNR (dB) and (b) gCNR measurements for 9-anechoic targets of the cyst phantom in Fig. 6.

*4.2. Phantom Study*

*4.2.1. Wire-target phantom*

To assess performance improvements in more realistic circumstances, an experiment was conducted using a wire-target phantom. Fig. 8 presents the B-mode results (with a dynamic range of 60 dB) reconstructed by DAS with 1 PW, DAS with 75-PWs, DMAS with 1-PW, DL-SP with 1-PW, DL-GAN with 1-PW and the proposed DL-DCL with 1-PW in the phantom study. In the visual assessment, DAS with 1-PW produced the lowest image quality among the six beamforming methods with high side-lobe artifacts, while DAS-75-PWs yielded the highest image quality with high spatial resolution and clear boundaries of the hyperechoic target, as shown in Figs. 8(a) and 8(b). As illustrated in Fig. 8(c), DMAS with 1-PW suffered from severe dark region artifacts, while the lateral resolution was improved. The DL-based method with a supervised learning framework (i.e., DL-SP) showed moderate performance with improved lateral resolution and low side-lobe artifacts, and the DL-based method based on the GAN architecture was worse in image quality than DL-SP in terms of lateral resolution and side-lobe suppression, as illustrated in Figs. 8(d) and 8(e). The proposed unsupervised learning-based DL method (i.e., DL-DCL) significantly improved both lateral resolution and side-lobe suppression with minimal dark region artifacts, as depicted in Fig. 8(f).

Lateral and axial FWHMs were also measured using the four wire targets indicated by the blue box in Fig. 8(a) for quantitative evaluation of spatial resolution, and they are listed in Tables 4 and 5. As listed in Table 4, DL-DCL and DMAS showed significant improvements that were comparable with DAS-75PWs in lateral resolution. DL-SP also produced better lateral resolution than DAS with 1-PW, but DL-GAN showed no improvements compared with it. For axial resolution, all six approaches presented similar performance (i.e., no improvements), as listed in Table 5.

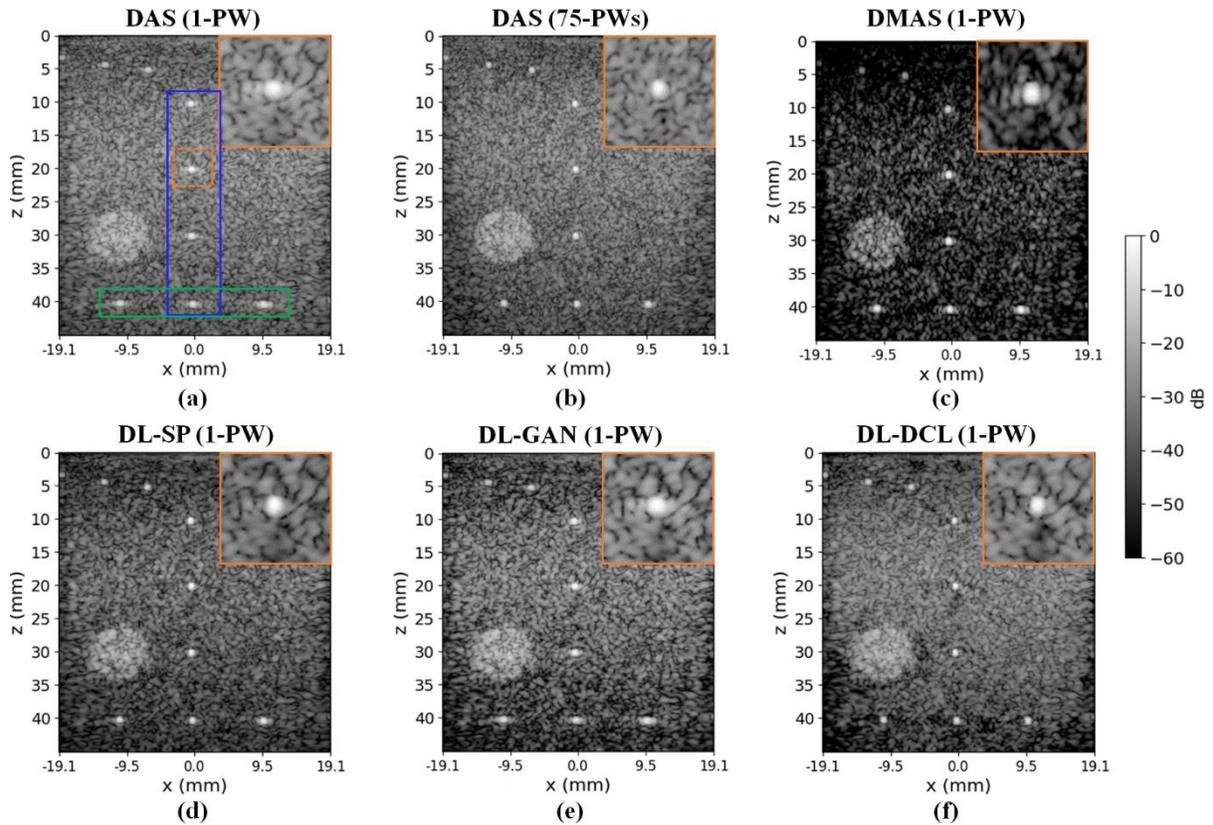

**Fig. 8.** B-mode imaging results reconstructed by (a) DAS with 1-PW, (b) DAS with 75-PWs, (c) DMAS with 1-PW, (d) DL-SP with 1-PW, (e) DL-GAN with 1-PW and (f) DL-DCL with 1-PW (proposed) from the wire-target phantom. The blue and green boxes in (a) indicate the wire targets for lateral and axial

FWHM measurements in Tables 4 and 5 and Table 7, respectively.

**Table 4.** The measured lateral FWHM (mm) for the four wire targets indicated by the blue box in Fig. 8(a).

|       | DAS (1-PW) | DAS (75-PWs) | DMAS (1-PW) | DL-SP (1-PW) | DL-GAN (1-PW) | DL-DCL (1-PW) |
|-------|------------|--------------|-------------|--------------|---------------|---------------|
| 10 mm | 0.59       | 0.39         | 0.39        | 0.39         | 0.59          | 0.39          |
| 20 mm | 0.59       | 0.39         | 0.39        | 0.49         | 0.59          | 0.39          |
| 30 mm | 0.59       | 0.39         | 0.39        | 0.39         | 0.59          | 0.39          |
| 40 mm | 0.59       | 0.39         | 0.39        | 0.49         | 0.69          | 0.39          |

**Table 5.** The measured axial FWHM (mm) for the four wire targets indicated by the blue box in Fig. 8(a).

|       | DAS (1-PW) | DAS (75-PWs) | DMAS (1-PW) | DL-SP (1-PW) | DL-GAN (1-PW) | DL-DCL (1-PW) |
|-------|------------|--------------|-------------|--------------|---------------|---------------|
| 10 mm | 0.49       | 0.49         | 0.39        | 0.49         | 0.49          | 0.49          |
| 20 mm | 0.49       | 0.49         | 0.49        | 0.49         | 0.49          | 0.49          |
| 30 mm | 0.49       | 0.39         | 0.49        | 0.39         | 0.49          | 0.49          |
| 40 mm | 0.49       | 0.49         | 0.49        | 0.49         | 0.49          | 0.49          |

*4.2.1. Cystic phantom*

A cystic phantom was also evaluated to investigate improvements in image contrast in real situations. Fig. 9 shows the results of B-mode images (with a dynamic range of 60 dB) reconstructed by the six beamforming methods (i.e., DAS with 1-PW, DAS with 75-PWs, DMAS with 1-PW, DL-SP with 1-PW, DL-GAN with 1-PW and DL-DCL with 1-PW). As illustrated in Figs. 9(b) and 9(f), DAS with 75-PWs and the proposed DL-DCL with 1-PW showed great distinction between the anechoic region and isoechoic background with improved contrast resolution. The DMAS method produced low contrast since the nonlinear operation led to signal distortion that resulted in increased speckle variance, as depicted in Fig. 9(c). The two DL-based methods (i.e., DL-SP and DL-GAN) showed insignificant improvements due to high remaining side-lobe artifacts compared to DAS-1-PW, as illustrated in Figs. 9(d) and 9(e).

Table 6 lists CNR and gCNR measurements using the anechoic and background regions, as indicated by Fig. 9(a). As presented in Table 6, DAS with 1-PW and DMAS with 1-PW showed the lowest values in both CNR and gCNR measurements, and DL-SP and DL-GAN exhibited better values than them. Interestingly, the proposed DL-DCL method showed a higher CNR value than DAS with 75-PWs (i.e., 16.31 vs. 14.58 dB), while the two methods possessed similar highest gCNR values.

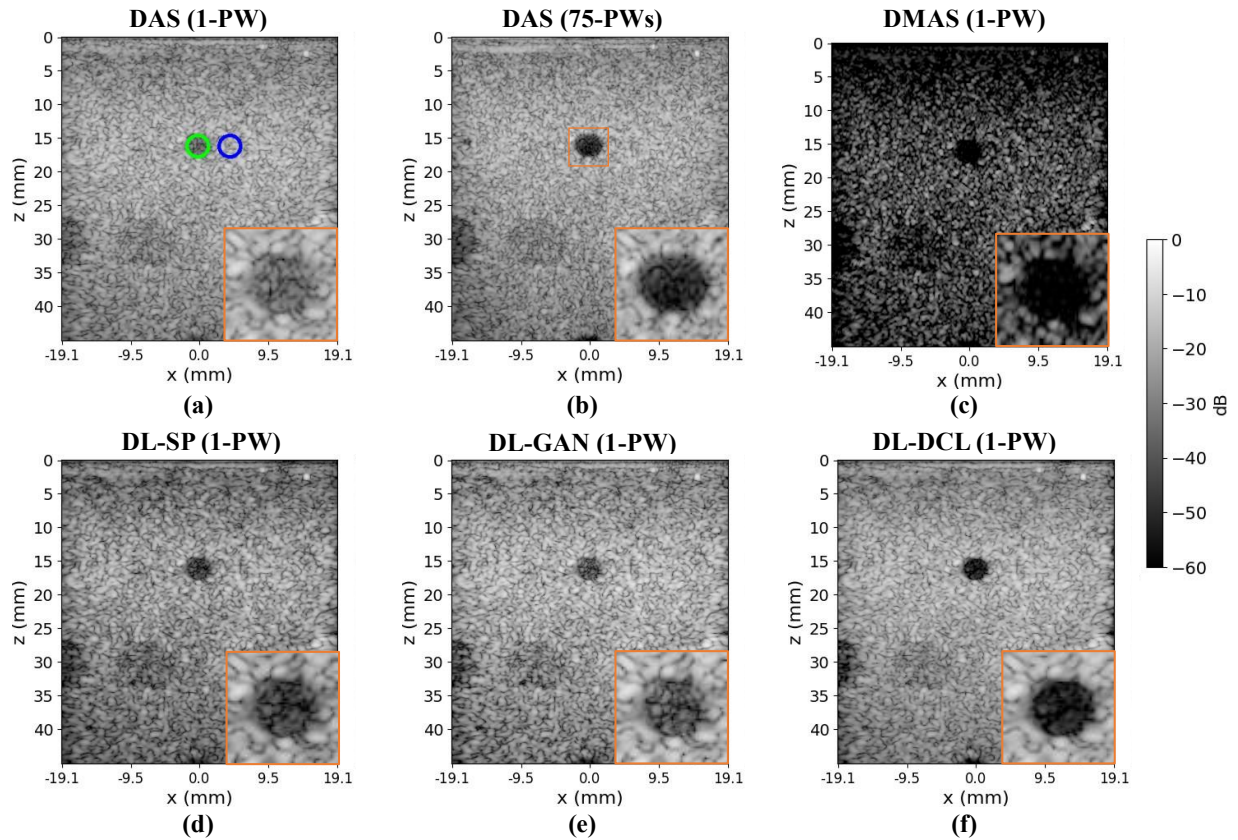

**Fig. 9.** B-mode imaging results reconstructed by (a) DAS with 1-PW, (b) DAS with 75-PWs, (c) DMAS with 1-PW, (d) DL-SP with 1-PW, (e) DL-GAN with 1-PW and (f) DL-DCL with 1-PW (proposed) using the cystic phantom. The green and blue circles in (a) indicate the anechoic and isoechoic background regions, respectively, for contrast measurements (i.e., CNR and gCNR).

**Table 6**. The measured CNR (dB) and gCNR for the cyst phantom.

| | DAS (1-PW) | DAS (75-PWs) | DMAS (1-PW) | DL-SP (1-PW) | DL-GAN (1-PW) | DL-DCL (1-PW) |
|---|---|---|---|---|---|---|
| CNR (dB) | 9.13 | 14.58 | 8.49 | 11.72 | 10.35 | 16.31 |
| gCNR | 0.87 | 0.99 | 0.90 | 0.96 | 0.92 | 0.99 |

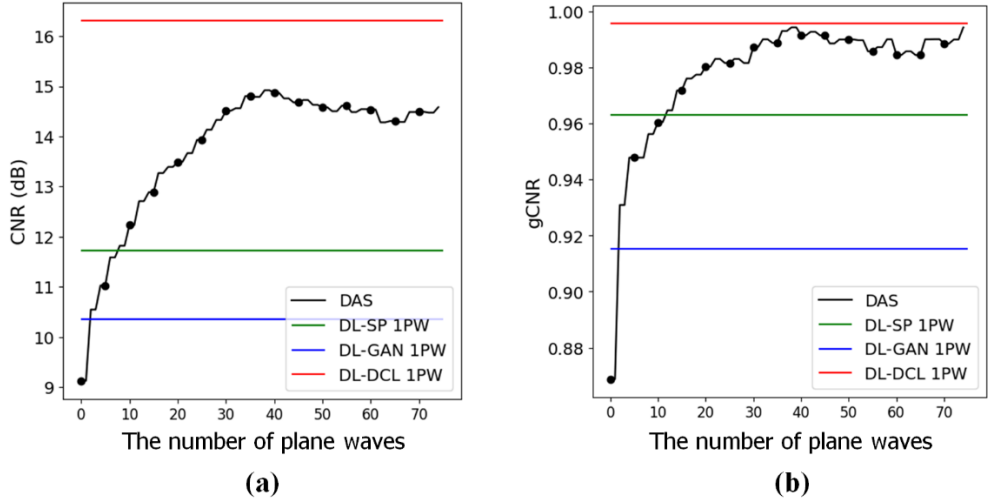

**Fig. 10.** (a) CNR (dB) and (b) gCNR comparisons between DAS with varying numbers of plane waves (i.e., CPWC) and the three DL-based methods (i.e., DL-SP, DL-GAN and DL-DCL) with 1-PW.

In addition, the three DL-based beamforming approaches (i.e., DL-SP, DL-GAN and DL-DCL) with 1-PW were compared to the traditional DAS beamformer with varying numbers of plane waves (i.e., CPWC). Fig. 10 illustrates the measured CNR and gCNR values using the same region-of-interest in Fig. 9. As shown in Fig. 10, both CNR and gCNR values of DAS beamforming increased as the number of plane waves increased. The DL-SP method showed better performance in both CNR and gCNR than DL-GAN, as already listed in Table 6. The proposed DL-DCL outperformed all DAS and exhibited higher CNR and gCNR values than DAS for all numbers of plane waves.

Furthermore, the studies that most recently introduced DL-based single PWI (Lu et al., 2022; Zhang et al., 2021) were also compared in terms of lateral and axial FWHMs and CNR since the same PICMUS dataset (i.e., Figs. 8 and 9) was utilized. As shown in Table 7, the proposed DL-DCL method outperformed the two recent studies in all aspects, and, specifically, DL-DCL greatly differed between them in terms of CNR (i.e., 16.3 vs. 11.7 and 8.3).

**Table 7.** Comparison with other studies (Lu et al., 2022; Zhang et al., 2021) in terms of lateral and axial FWHMs and CNR (dB) using the PICMUS dataset (i.e., Figs. 8 and 9).

|  | DAS (1-PW) | DAS (75-PWs) | Zhang et al. | Lu et al. | DL-DCL |
|---|---|---|---|---|---|
| Lateral FWHM (mm) | 0.49 | 0.47 | 0.52 | 1.04 | 0.43 |
| Axial FWHM (mm) | 0.49 | 0.47 | 0.52 | 1.02 | 0.52 |
| CNR (dB) | 9.1 | 14.6 | 11.7 | 8.3 | 16.3 |

*4.3. In vivo Study*

A feasibility study using an *in vivo* dataset was performed, and the results were analyzed in a similar manner. Fig. 11 illustrates the reconstructed B-mode images using DAS with 1-PW, DAS with 75-PWs, DMAS with 1-PW, DL-SP with 1-PW, DL-GAN with 1-PW and the proposed DL-DCL with 1-PW for a common carotid artery with a longitudinal view. The images were displayed with a dynamic range of 60 dB. As shown in Fig. 11, conventional DAS beamforming with 1-PW impaired the image quality (e.g., contrast) by secondary PW artifacts, and it was difficult to distinguish boundaries between the vessel lumen and background tissues. DAS with 75-PWs depicted anatomical structures (i.e., lumen and background tissues) clearly with high spatial resolution and contrast, while diffusive reverberation artifacts (the white arrow in Fig. 11(b)) remained, as illustrated in Fig. 11(b). DMAS nonlinear beamforming with 1-PW, as shown in Fig. 11(c), presented clear boundaries of the vessel wall, but the overall image quality was severely deteriorated by dark region artifacts. As illustrated in Figs. 11(d) and 11(e), the DL-SP and DL-GAN methods with 1-PW exhibited better image quality than DAS with 1-PW from moderate image contrast. In the unsupervised DCL-based DL method, image contrast was significantly increased with clear boundaries and structures, and the diffusive reverberation artifacts were most effectively suppressed among the six comparison methods, as shown in Fig. 11(f).

Table 8 lists the measured CNR and gCNR using anechoic (vessel lumen) and hyperechoic (vessel wall) regions as indicated by the green and blue boxes in Fig. 11(a). As listed in Table 8, DAS with 75-PWs highly increased both CNR and gCNR values compared to DAS with 1-PW. DMAS with the 1-PW method showed lower CNR and gCNR values than DAS with the 1-PW method, which was consistent with the visual assessment. DL-SP with 1-PW outperformed DAS with 1-PW and DMAS with 1-PW, and DL-GAN with 1-PW exhibited similar performance to DAS with 75-PWs. The proposed DL-DCL with 1-PW showed higher CNR and gCNR values than DAS with 75-PWs as expected in Fig. 11.

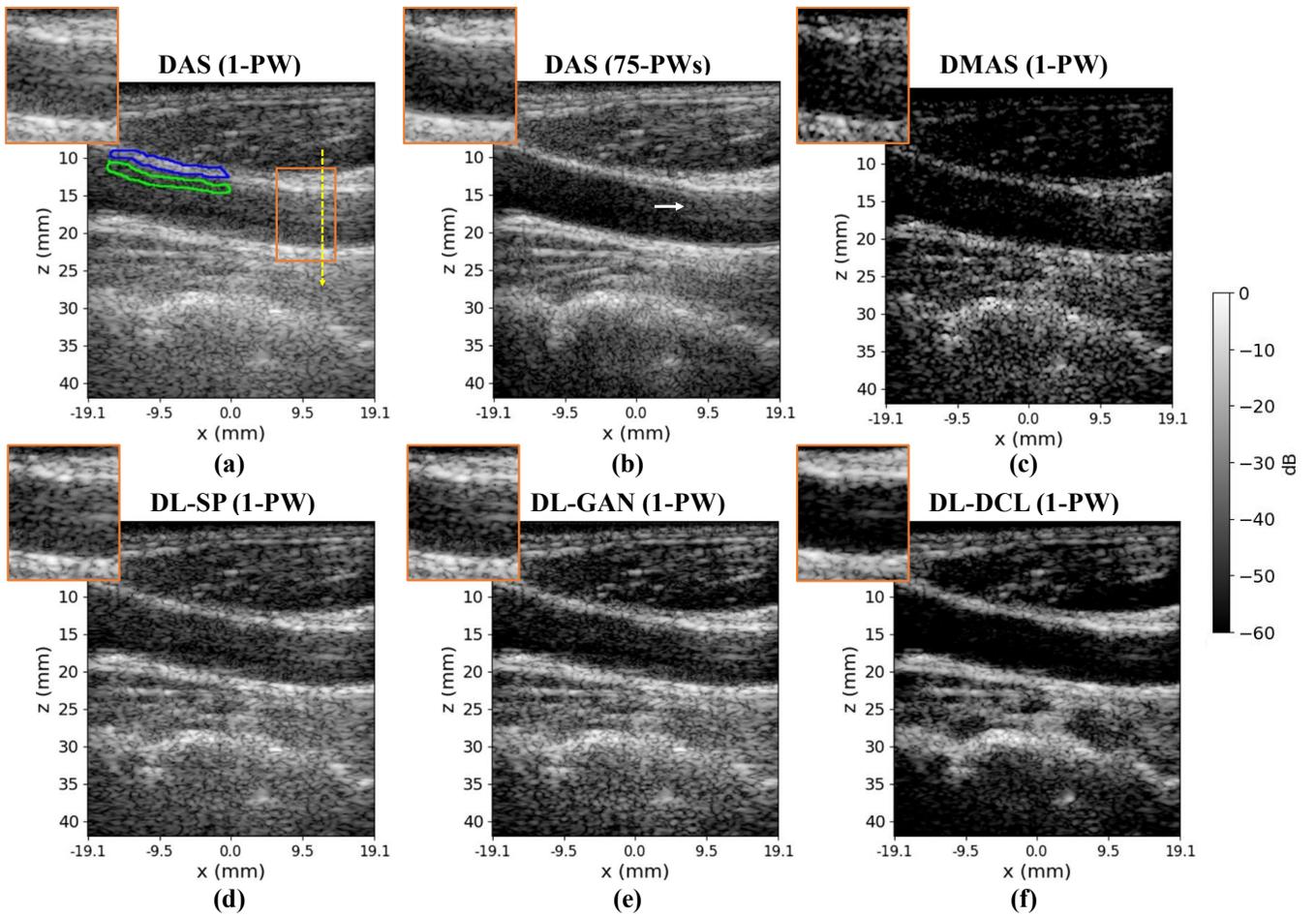

**Fig. 11.** *In vivo* B-mode results reconstructed by (a) DAS with 1-PW, (b) DAS with 75-PWs, (c) DMAS with 1-PW, (d) DL-SP with 1-PW, (e) DL-GAN with 1-PW and (f) DL-DCL with 1-PW (proposed) for the common carotid artery (longitudinal view). The green and blue contours in (a) indicate anechoic (vessel lumen) and hyperechoic (wall) regions for CNR and gCNR measurements. The yellow vertical dashed line also indicates the region for axial profiles in Fig. 13.

**Table 8.** The measured CNR (dB) and gCNR from *in vivo* carotid artery images in Fig. 11.

|          | DAS (1-PW) | DAS (75-PWs) | DMAS (1-PW) | DL-SP (1-PW) | DL-GAN (1-PW) | DL-DCL (1-PW) |
|----------|------------|--------------|-------------|--------------|---------------|---------------|
| CNR (dB) | 8.51       | 10.09        | 5.56        | 9.09         | 10.28         | 12.99         |
| gCNR     | 0.84       | 0.91         | 0.70        | 0.86         | 0.88          | 0.98          |

Fig. 12 exhibits another example of the *in vivo* results in the breast, including the mass area (i.e., fibroadenoma). Unlike the previous result in the carotid artery scan (i.e., Fig. 11), the image contrast of DAS with 75-PWs was not highly increased compared to DAS with 1-PW, while boundaries and structures were clearly depicted, as illustrated in Figs. 12(a) and 12(b). DMAS with 1-PW suffered from overall dark regions while the spatial resolution increased, as illustrated in Fig. 12(c), and DL-GAN with 1-PW showed better image contrast than DL-SP with 1-PW, as illustrated in Figs. 12(d) and 12(e). The DL-DCL method outperformed DAS-75-PWs in terms of image contrast in the visual assessment and resulted in better conspicuity of breast masses from the background.

Table 9 lists the measured CNR and gCNR using hypoechoic (mass) and hyperechoic (background) regions as indicated by the green and blue boxes in Fig. 12(a). The tendency of the results was consistent with the previous case (i.e., Table 8). The CNR and gCNR values of DAS with 1-PW were similar to those of DL-SP with 1-PW, and the values of DL-GAN with 1-PW were greater than those of the two methods (i.e., DAS with 1-PW and DL-SP with 1-PW). DMAS with 1-PW showed the lowest CNR and gCNR values among the six beamforming methods. DAS with 75-PWs yielded the second highest values in terms of CNR and gCNR, and DL-DCL with 1-PW outperformed DAS with 75-PWs while DL-DCL with 1-PW has a frame rate 75 times faster than DAS with 75-PWs.

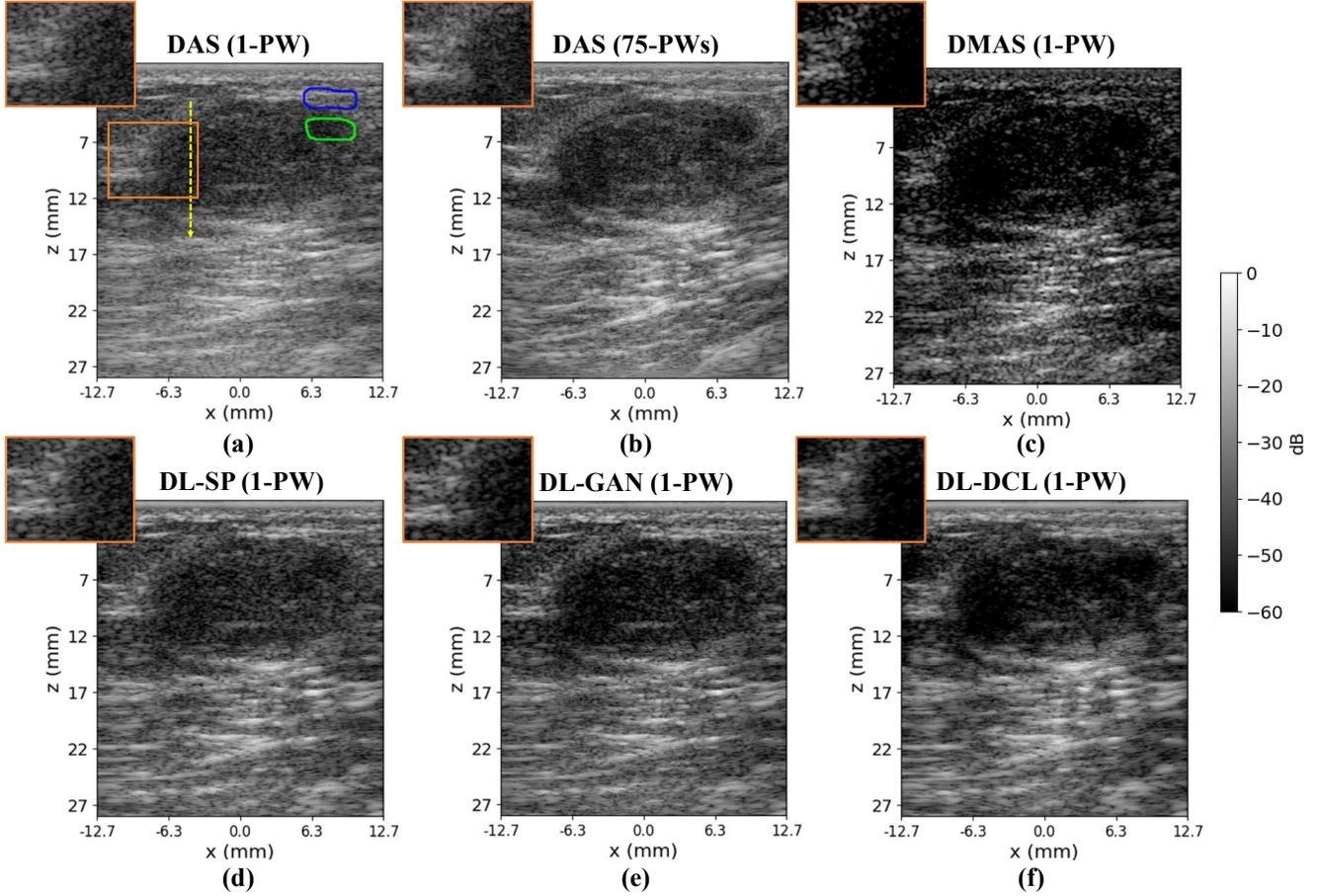

**Fig. 12.** *In vivo* B-mode results reconstructed by (a) DAS with 1-PW, (b) DAS with 75-PWs, (c) DMAS with 1-PW, (d) DL-SP with 1-PW, (e) DL-GAN with 1-PW and (f) DL-DCL with 1-PW (proposed) for breast mass (fibroadenoma). The green and blue contours in (a) indicate hypoechoic (mass) and hyperechoic (background) regions for CNR and gCNR measurements. The yellow dashed line indicates the region for axial profiles in Fig. 13.

**Table 9.** The measured CNR (dB) and gCNR from *in vivo* breast images in Fig. 12.

|          | DAS (1-PW) | DAS (75-PWs) | DMAS (1-PW) | DL-SP (1-PW) | DL-GAN (1-PW) | DL-DCL (1-PW) |
|----------|------------|--------------|-------------|--------------|---------------|---------------|
| CNR (dB) | 7.28       | 8.36         | 2.88        | 7.25         | 8.07          | 9.80          |
| gCNR     | 0.77       | 0.82         | 0.58        | 0.78         | 0.82          | 0.90          |

Figs. 13(a) and 13(b) represent the normalized axial profiles for the vessel lumen and breast mass regions, as indicated by the yellow dashed line in Fig. 11 and Fig. 12, respectively. As depicted in Fig.

13(a), diffusive reverberation artifacts were dominant in the lumen region, as indicated by the arrow, and DAS with 75-PWs and DL-GAN with 1-PW methods exhibited lower diffusive reverberation artifacts by approximately 6 dB than DAS with 1-PW, DMAS with 1-PW and DL-SP with 1-PW methods. On the other hand, the proposed DL-DCL method significantly suppressed the artifacts, while DAS with 75-PWs still suffers from clutter artifacts. For the axial profiles in the breast, as illustrated in Fig. 12(b), the DL-DCL method clearly distinguished the mass area (anechoic region) from the background, and it showed an approximately 10 dB maximal difference compared to DAS-75 PWs.

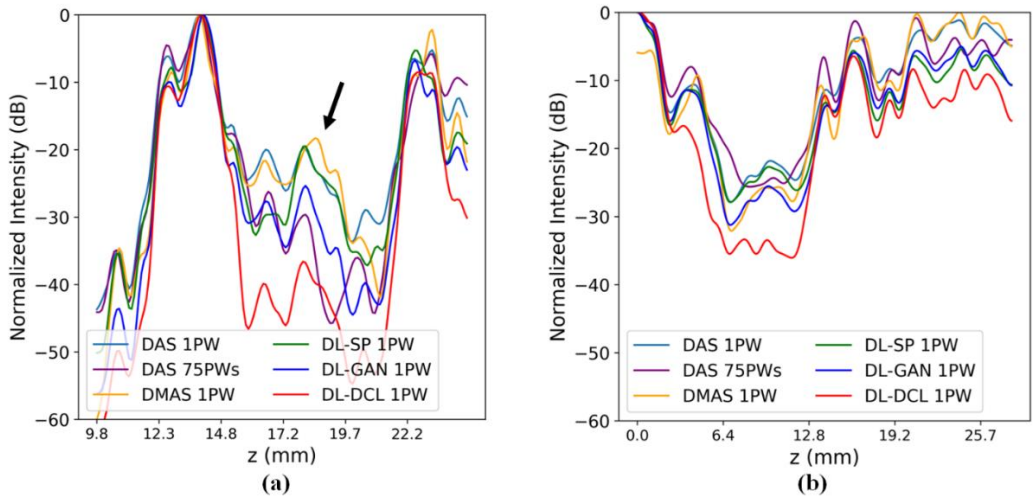

**Fig. 13.** Normalized axial profiles of the six images from Figs. 11 (carotid artery) and 12 (breast).

## 5. Discussion

*5.1. Benefits of DL-DCL*

Although single PWI in medical ultrasound can provide an extremely high frame rate compared to conventional focused beam imaging or CPWC imaging, the image quality is still challenging. In recent studies (Goudarzi and Rivaz, 2022; Khan et al., 2020; Li et al., 2020; Lu et al., 2022; Luijten et al., 2020; Nair et al., 2019; Nguon et al., 2022; Tang et al., 2021; Wasih et al., 2023; Zhang et al., 2021), deep learning-based adaptive beamformers have shown great potential for high-image quality single PWIs, and they have typically adopted a supervised learning approach in which neural networks are trained by ground-truth data generated by CPWC or adaptive beamforming techniques. However, due to the constraints of ground truth imposed by the nature of medical ultrasound, supervised learning-based approaches may not improve performance while minimizing additional artifacts.

In this article, a novel unsupervised learning-based beamforming approach, which is called deep coherence learning (DCL), was proposed for high-quality single PWI. As shown in Fig. 1, the proposed DCL identifies signal coherence between PWs, and it is reinforced by training with efficient DL networks. This means that DCL encourages DL networks to predict correlated signals from different PSFs of steered PWs rather than directly minimizing the difference between prediction and ground-truth data. Thus, this unsupervised learning approach enables DL networks to overcome limitations stemming from the current ground truth. To evaluate the effectiveness of the proposed DCL method, simulation, phantom and *in vivo* studies were conducted.

In the simulation study with single PW data, the DMAS nonlinear beamformer exhibited the most favorable performance in the improvement of lateral resolution, but it also showed severe dark regions as well as remaining axial-lobe artifacts. In contrast, general DL-based beamformers (i.e., DL-SP and DL-GAN) were more effective in axial-lobe suppression, but they did not successfully reduce side lobes and grating lobes. The proposed DL-DCL method significantly improved the lateral resolution and suppressed all secondary lobes (i.e., side lobes, grating lobes and axial lobes). Therefore, DL-DCL can achieve remarkably improved CNR and gCNR, which are also comparable to DAS with 75-PWs (reference image).

The improvement in the performance of DL-DCL was more obvious in the phantom and *in vivo* studies. For example, in the study with a cyst phantom (i.e., Figs. 9 and 10), the improvement in image contrast in conventional DAS beamforming was saturated despite the number of plane waves continuously increasing (meaning that the frame rate continuously decreases). Thus, DL-DCL with 1-PW outperformed DAS with 75-PWs in terms of both CNR and gCNR. On the other hand, a supervised learning-based DL method (i.e., DL-SP) using the ground truth of DAS with 75-PWs showed a comparable performance with DAS with 15-PWs, while it had better performance than DL-GAN. In addition, the DL-based beamformers yielded low dark region artifacts, while the DMAS adaptive beamformer led to highly increased speckle variance by nonlinear processing, as illustrated in Figs. 8 and 9. The consecutive results were shown in the *in vivo* study, as presented in Figs. 11 and 12 and Tables 8 and 9, and DL-DCL effectively improved spatial and contrast resolution while minimizing dark region artifacts. As a result, *in vivo* images were significantly improved by identifying useful clinical information more clearly.

*5.2. Limitations and Future Works*

In spite of the benefits of the DL-DCL method, it has several limitations. In the point target study, as illustrated in Fig. 4(f), the performance of the DL-DCL method was degraded in the near depth (~10 mm) since it retained relatively higher side-lobe and grating-lobe artifacts compared to the deeper targets. Those artifacts were raised due to less correlated signals from the limited transmit synthetic fields in CPWC. This may be addressed by preprocessing the input dataset where dynamic weighting factors (apodization) are applied to signals in the near depth.

In the real phantom and *in vivo* experiments, some dark region artifacts occurred around hyperechoic targets, although they were minimized and compromised for improvements in spatial resolution and secondary artifact removal, as illustrated in Figs. 8(f) and 11(f). This phenomenon is generally observed in traditional adaptive beamforming techniques such as DMAS, as shown in Figs. 8(c) and 11(c). Future studies should focus on this issue, and an enhanced loss function based on signal coherence may be proposed.

In addition, an improved DL network for the unsupervised DCL will be newly designed since a typical encoder-decoder-based CNN structure (Fig. 3) was adopted in this study. For this network, a new network structure to mitigate the limitations presented above will be studied. In terms of implementing a DL-driven single PWI system with an exceptionally high frame rate, the inference speed of DL-DCL should be more accelerated. For this, the recently proposed network compression techniques (Han et al., 2015; Marinó et al., 2023) will also be investigated. On the other hand, DL-DCL might have potential for PWI applications, such as shear wave elastography or pulse-wave velocity imaging (Kang et al., 2022; Taljanovic et al., 2017), which requires phase-sensitive data with extremely high frame rates.

## 6. Conclusions

In this article, a new unsupervised learning approach for high-quality images of a single PWI, named DCL, was proposed. The DCL leads the DL model to predict highly correlated signals from the CPWC dataset by a unique loss function, and the trained DL model encourages a high-image quality PWI from low-quality single PW data. To evaluate performance, simulation, phantom and *in vivo* studies were conducted, and the results were compared with those from several adaptive beamformers based on nonlinear operation (i.e., DMAS) and other DL networks (i.e., supervised learning method and GAN structure) as well as DAS with 75-PWs. In experiments, the proposed DCL showed comparable results with DMAS and DAS with 75-PWs in terms of spatial resolution, and it outperformed all comparison methods in terms of contrast resolution. The feasibility studies indicated that the proposed unsupervised learning approach can overcome the limitations of traditional DL methods based on supervised learning and showed great potential in clinical settings with minimal artifacts.

## Acknowledgments

This work was supported in part by the Korea Medical Device Development Fund grant funded by the Korea government (the Ministry of Science and ICT, the Ministry of Trade, Industry and Energy, the Ministry of Health & Welfare, Republic of Korea, the Ministry of Food and Drug Safety) (Project Number: 202011A01) and in part by Basic Science Research Program through the National Research Foundation of Korea (NRF) funded by the Ministry of Education (Project Number: 2021R1A6A3A14039529).